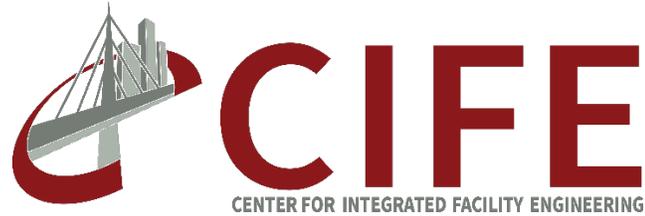

# The Application of Blockchain-Based Crypto Assets for Integrating the Physical and Financial Supply Chains in the Construction & Engineering Industry

By

## Hesam Hamledari & Martin Fischer

CIFE Technical Report #TR245
December 2020
**STANFORD UNIVERSITY**

# The Application of Blockchain-Based Crypto Assets for Integrating the Physical and Financial Supply Chains in the Construction & Engineering Industry

**Hesam Hamledari[1], Martin Fischer[2]**

[1] PhD Candidate, Stanford University, Department of Civil & Environmental Engineering, Stanford, CA, United States, hesamh@stanford.edu

[2] Kumagai Professor of Engineering and Professor in Civil & Environmental Engineering, Stanford University, Stanford, CA, United States, fischer@stanford.edu

## ABSTRACT

Supply chain integration remains an elusive goal for the construction and engineering industry. The high degree of fragmentation and the reliance on third-party financial institutions has pushed the physical and financial supply chains apart. The paper demonstrates how blockchain-based crypto assets (crypto currencies and crypto tokens) can address this limitation when used for conditioning the flow of funds based on the flow of products. The paper contrasts the integration between cash and product flows in supply chains that rely on fiat currencies and crypto assets for their payment settlement. Two facets of crypto asset-enabled integration, atomicity and granularity, are further introduced. The thesis is validated in the context of construction progress payments. The as-built data captured by unmanned aerial and ground vehicles was passed to an autonomous smart contract-based method that utilizes crypto-currencies and crypto tokens for payment settlement; the resulting payment datasets, written to the Ethereum blockchain, were analyzed in terms of their integration of product and cash flow. The work is concluded with a discussion of findings and their implications for the industry.

# 1 INTRODUCTION

The construction and engineering industry has long been in the pursuit of supply chain integration (Howard et al. 1989; O'Brien and Fischer 1993). Most efforts have focused on increasing the strategic collaboration and partnership between the construction supply chain partners (Briscoe and Dainty 2005; Cheng et al. 2010; Dainty et al. 2001). A goal that is equally important and often neglected is the integration of flows (Rai et al. 2006) and in particular the cash and product flows (Blount 2008; Hofmann 2005). In an integrated supply chain, the flow of products (physical supply chain) and funds (financial supply chain) need to be integrated throughout the life cycle of a project (Rai et al. 2006); they should not be separate. Integrating the physical and financial supply chain enables supply chain optimization (Kifokeris and Koch 2020) and creates a single source of truth, key to successful delivery of projects (Barbosa 2017; Fischer et al. 2017).

Despite the desired benefits of such integration, the physical and financial supply chains have been pushed apart due to the high degree of fragmentation in the industry (Bilal et al. 2016; Maria João Ribeirinho et al. 2020). This fragmentation stems from the high number of stakeholders (Briscoe and Dainty 2005; Cheng et al. 2010) and the siloed data development in the industry's information technology infrastructure (Barua et al. 2004; Chen et al. 2018).

To manage its financial supply chain, the fragmented construction industry has to rely on trusted third parties, banks and financial institutions (Bitran et al. 2007; Feldmann and Müller 2003). These institutions play a role that is key to achieving integration (Blount 2008), yet their involvement as an external stakeholder has hurt the integration between the physical and financial supply chains (Blount 2008; Fellenz et al. 2009; John Mathis and Cavinato 2010). This lack of integration has led to misalignments between the documentations of flows across different project information sources (Čuš-Babič et al. 2014).

This paper demonstrates how blockchain-based crypto assets can address this limitation when used for conditioning payments based on the flow of products. The underlying mechanisms contributing to crypto asset-enabled integration are elaborated by contrasting the flow of cash in today's supply chains, reliant on fiat currency, and those using crypto assets. Two facets of crypto asset-enabled integration, atomicity and granularity, are further introduced. To validate this work's thesis, a series of experiments are conducted where crypto assets are used for payment processing and lien right management on two commercial construction projects. The robot-captured observations of two job sites are passed to a smart contract-based method that uses crypto currencies and crypto tokens for processing payments to subcontractors. The resulting payment datasets, written to a public blockchain, are examined in terms of their integration of cash and product flow. The paper is concluded with a discussion of findings, limitations, and the implications for the AEC industry.

## 2    POINT OF DEPARTURE

Blockchain and smart contract are the underlying technologies that empower crypto assets. This section first reviews the studies around the use of these two technologies in managing cash flow and its integration with product flow (section 2.1). This is followed by a review of crypto asset classes (section 2.2). In this work, the terms "financial supply chain", "cash flow", "the flow of cash", and "the flow of funds" are used interchangeably.

### 2.1    Applications of Smart Contract for Cash Flow Management

Cash flow management is currently governed by inefficient contracts that rely on manual workflows and third party interference (Salleh et al. 2020); these limitations make traditional contracts a major source of delay (Odeh and Battaineh 2002), distrust (Gabert and Grönlund 2018; Manu et al. 2015), and information asymmetry between supply chain partners (Xiang et al. 2015).

Blockchain-enabled smart contracts are believed to provide cost and time savings by reducing the administrative work (Sreckovic and Windsperger 2019) and to promote collaboration by increasing transparency (McNamara and Sepasgozar 2018). Therefore, smart contracts can reduce the cost of contracting (Qian and Papadonikolaki 2020). Others argue that smart contracts' true value proposition in cash flow management is twofold: 1) they increase the *confidence* in the output of the computational systems (De Filippi et al. 2020) and the input data (Penzes et al. 2018) used in payment processing; and 2) they add *reliability* to payment automation due to their elimination of centralized control mechanisms and providing guarantee of execution (Hamledari and Fischer 2020a). While alternative technologies may support automation, they lack such reliability. An example is internet-based payment applications; they are proven to decrease the processing time by 84% compared with paper-based payment applications (Barrón and Fischer 2001), but they still rely on the same workflows that hurt traditional contracts.

Despite these potential benefits, practitioners have doubts about the technology's capabilities (Mason and Escott 2018). The adoption will remain slow (Sharma and Kumar 2020) until the industry adapts its policies (Hamma-adama et al. 2020; Li et al. 2019), addresses the legal hurdles (Badi et al. 2020), and develops new business models (Tezel et al. 2020). Others argued that semi-automated contracts (Altay and Motawa 2020) and modular construction (Owusu et al. 2020) provide a more feasible path to adoption due to their relatively simpler contract structure.

One study (Elghaish et al. 2020) proposed a framework for the execution of financial transactions in integrated project delivery (IPD) projects using smart contracts. The method, based on Hyperledger Fabric (Androulaki et al. 2018), keeps track of achieved profit in comparison with planned profit. Others argued that the alignment between IPD and blockchain's incentive structure can enhance collaboration (Hunhevicz et al. 2020).

To improve the security of payments, a method was proposed (Ahmadisheykhsarmast and Sonmez 2020) to lock funds in a smart contract account for a period of 30 days, reducing the trades' exposure to the insolvency of clients. To enable the sharing of payment records at project-level, a key management strategy was proposed (Das et al. 2020) to provide "selective-transparency", keeping sensitive information only visible between two contracting parties. A semi-autonomous solution (Luo et al. 2019) proposed the use of a consensus mechanism in the place of today's payment workflows; contractors submit data regarding the applications for payment which is processed in a decentralized peer-to-peer network of stakeholders. Payments are executed by a smart contract if the relevant stakeholders reach consensus.

The integrations with BIM is key to successful applications of the smart contracts (Mason 2017), and this has been the focus of recent work on payment automation. A framework was introduced for using 5D BIM in the context of automated billing (Ye and König 2020; Ye et al. 2020), extracting the bill of quantities from the project models and enhancing transparency; the framework needs to be implemented, validated, and imporve its design to include payments to subcontractors.

Two studies took a more integrated approach, focusing on the management of the cash flow in relationship with the physical supply chain: a framework was proposed (Chong and Diamantopoulos 2020) for conditioning the smart contract's payments based on the sensor feed and the BIM used for tracking the on-site installation of building façade panels. This work created a link between off- and on-chain realities by its direct use of product flow for payment processing. The smart contract design was not detailed however. A smart contract-enabled solution (Hamledari and Fischer 2020b) was introduced to autonomously translate the on-site reality captures to direct payments to general contractor and subcontractors; it eliminated the need for payment applications. The progress data and as-built BIMs are stored off-chain on a private InterPlantary File Sharing (IPFS) (Benet 2014) with their cryptographic summary stored on the Ethereum blockchain (Buterin 2014) and used for on-chain payment settlement.

It was argued (Hamledari and Fischer 2020a) that such autonomous conditioning of cash flow on product flow, as described in the two studies above, cannot be achieved with other technological alternatives. The Cash and product flows are respectively institutional and brute facts; the former is a social reality, relying on the collective agreement of stakeholders. The transition from brute to institutional facts necessitates an elimination of single points of failure and centralized control mechanisms that are present in payment applications; smart contracts make this possible (Hamledari and Fischer 2020a). A conceptual blockchain-based business model (Kifokeris and Koch 2020) was introduced for use by construction logistics consultants and in support of integrating supply chain flows. The process flows and the the logistics set up was motivated through a review of literature and empirical findings in the context of sweedish construction industry.

The review of the literature reveals a need for increased attention on the integration between physical and financial supply chains (Kifokeris and Koch 2019). The research landscape has remained mostly theoretical (Darabseh and Martins 2020; Hunhevicz and Hall 2020; Kasten 2020) and lacks validation of usability (Hijazi et al. 2019). While a handful of studies focused on the transition from product flow to cash flow, it is not clear whether this transition leads to integration and how such integration compares with that of today's construction supply chain. In addressing this challenge, this work focuses on the role of crypto assets as enablers of integration between physical and supply chain. The section 2.2 provides a brief overview of crypto asset classes, a point of departure used in this work.

## 2.2 Crypto asset

The term *"crypto asset"* is used herein to refer to both crypto currencies and crypto tokens. At their core, crypto assets provide a decentralized governance model for reaching agreement on a shared notion of reality in trust-less environments.

### 2.2.1 Crypto Currency

In the case of crypto currencies, this shared notion is the concept of money and its exchange between a network of peers. Money is an institutional fact (Searle 1995), and it requires the active participation of trusted third-party institutions such as banks who define and back this shared notion of value. The invention of Bitcoin (Nakamoto 2008) and its Nakamoto consensus allowed for independent parties, with competing objectives, to collectively execute and agree on the exchanges of monetary value (bitcoin) without reliance on trusted third-party intermediaries.

In their early writings, the Bitcoin inventor Satoshi Nakamoto referred to the blockchain as "proof-of-work chain" (Champagne 2014). This further emphasizes the critical role of the *consensus/governance model* as the core contribution of Bitcoin (Tschorsch and Scheuermann 2016), and not the use of a shared distributed ledger. This decentralization reduces the cost of transaction verification and networking (Catalini and Gans 2016), and it is the innovation that distinguishes Bitcoin and other crypto currencies from their unsuccessful predecessors (Narayanan et al. 2016; Narayanan and Clark 2017) such as B-Money (Dai 1998), and Bit Gold (Szabo 2005).

Bitcoin's shortcomings gave rise to alternative coins ("alt-coin") (Antonopoulos 2014). The Litecoin (Lee 2011) was announced as a "lite version of Bitcoin", and it replaced the Bitcoin's SHA-256 with the Scrypt algorithm. This made its mining more accessible since the Scrypt algorithm is less susceptible to custom-hardware solutions and the resulting high consolidation of miners. Privacy is a concern due to the Bitcoin's use of pseudonymous transactions (Conti et al. 2018). The study of transaction patterns may jeopardize users' spending habits and their identity. For example, one study successfully distilled millions of transactions into a few thousand superclusters, each representing a business entity (Tasca et al. 2018). This need for privacy gave rise to Monero (Van Saberhagen 2013), Dash (Duffield and Diaz 2015), and Zerocash (also Zcash) (Sasson et al. 2014). For example, Zcash provides zero-knowledge proofs for transactions without revealing their source, amount, and destination. This is achieved with the combined use of an anonymous and a base non-anonymous currency. In addition to privacy, there are concerns with respect to scalability and the limited transaction throughput (Eyal et al. 2016).

While both permission-less and private chains use shared distributed legers to manage the exchanges of their native coins, their functionality and security stem from the design of their consensus mechanism (Antonopoulos 2017a). The mere use of a distributed database for exchange of crypto currencies does not guarantee that transactions are irreversible, and that consensus can be emerged. The anti-trust risks associated with consortium blockchains (Schrepel 2019; Schrepel and Buterin 2020) are one example of such challenges; this can posit difficulties for distributed leger technology (DLT)-based solutions (Mills et al. 2016).

While alt-coins face difficulties in bootstrapping (Böhme et al. 2015), they continue to gain adoption and constitute a bigger portion of the market. As a result, the support for Bitcoin has slightly declined, dropping from 98% in 2017 to 90% in 2020 (Blandin et al. 2020). Regardless of which crypto currencies survive, their innovation in the decentralized exchange of money is here to stay (Lo and Wang 2014).

### 2.2.2 Crypto Token

The invention of Bitcoin made *decentralized money* a reality, yet it fell short of supporting *decentralized applications* (DApp) due to its limited stack-based scripting language and its lack of turning completeness (Swan 2015). This motivated the invention of the Ethereum blockchain (Buterin 2014; Wood 2014) and its quasi-Turning complete Ethereum Virtual Machine (EVM). Ether (ETH) is the native coin of the Ethereum blockchain. This innovation enabled computerized algorithms to be executed in a decentralized manner, making earlier visions such as smart contract (Szabo 1994; Szabo 1997) a reality.

*Crypto tokens* are one example of smart contracts. A crypto token contract is executed on the EVM, and it manages its supply of tokens and their exchanges by updating a set of variables written to the underlying Ethereum blockchain. For smart contracts to transact with one another and for trading to take place, these token contracts and their functionalities needed to be standardized. This gave birth to the Ethereum Request for Comments (ERC) documents that describe a token standard. The two notable token standards are the ERC20 standard (Vogelsteller and Buterin 2015) and ERC721 standard (Entriken et al. 2018), respectively used to represent *fungible* and *non-fungible* digital assets.

ERC20 tokens are fungible; each token can be swapped with another. In contrast to fungible crypto currencies, however, these tokens can have additional functionalities, acting as "programmable money" (Antonopoulos 2017b). For example, they can represent a share in a venture; this has given rise to initial coin offerings (ICO), the crypto world's parallel to initial public offerings and a means of fund raising (Howell et al. 2020). The ERC20's popularity and utilization has made ETH the second most common crypto currency (Rauchs et al. 2018) trailing bitcoin.

ERC721 tokens are unique and cannot be swapped with one another. This enables tokenization, with each non-fungible token (NFT) representing a unique asset. This concept, however, has origins in the earlier writings by Nick Szabo on the decentralized exchanges of secure property titles (Szabo 1998) and the efforts in the Bitcoin ecosystem for the development of "colored coins" (Rosenfeld 2012). The application of NFTs has been proposed for infrastructure financing (Tian et al. 2020), the lien right management in construction industry (Hamledari and Fischer 2020b), life cycle management of information in AEC (Succar and Poirier 2020), and tracking goods across supply chains (Westerkamp et al. 2020).

## 3 CRYPTO ASSET-ENABLED INTEGRATION OF PHYSICAL AND FINANCIAL SUPPLY CHAINS

This work argues that the application of crypto assets in the place of fiat currencies and for conditioning payments on the updates in the product flow status can increase the integration between physical and financial supply chains. The authors first elaborate the underlying mechanisms that create such integration by contrasting the flow of cash in supply chains that are rely on fiat currencies and those using crypto assets (section 3.1). Two types of integration enabled by crypto assets are further introduced in section 3.2.

### 3.1 The Movement of Money: Seeming Versus Actual

Financial institutions such as banks have become indispensable parts of today's construction supply chain (Fig 1), responsible for processing payments in fiat currencies throughout a project's life cycle and between all its stakeholders. Banks operate based on the concept of *liability*. An account balance represents the sum *owed* to an account holder by a bank, and not reserves of cash. Banks process payments by adjusting their liability and the amounts they owe to project stakeholders on the receiving and sending end of a payment. A bank owes respectively more and less to the payee and payer; these changes in liability are respectively called credit and debit operations. This process is more straightforward when both stakeholders hold accounts with the same bank (Fig. 1b), where the financial institution debits one account by the exact amount it credits the other. The bank sees no change in its net liabilities, but instead changes in its liability to each individual stakeholder.

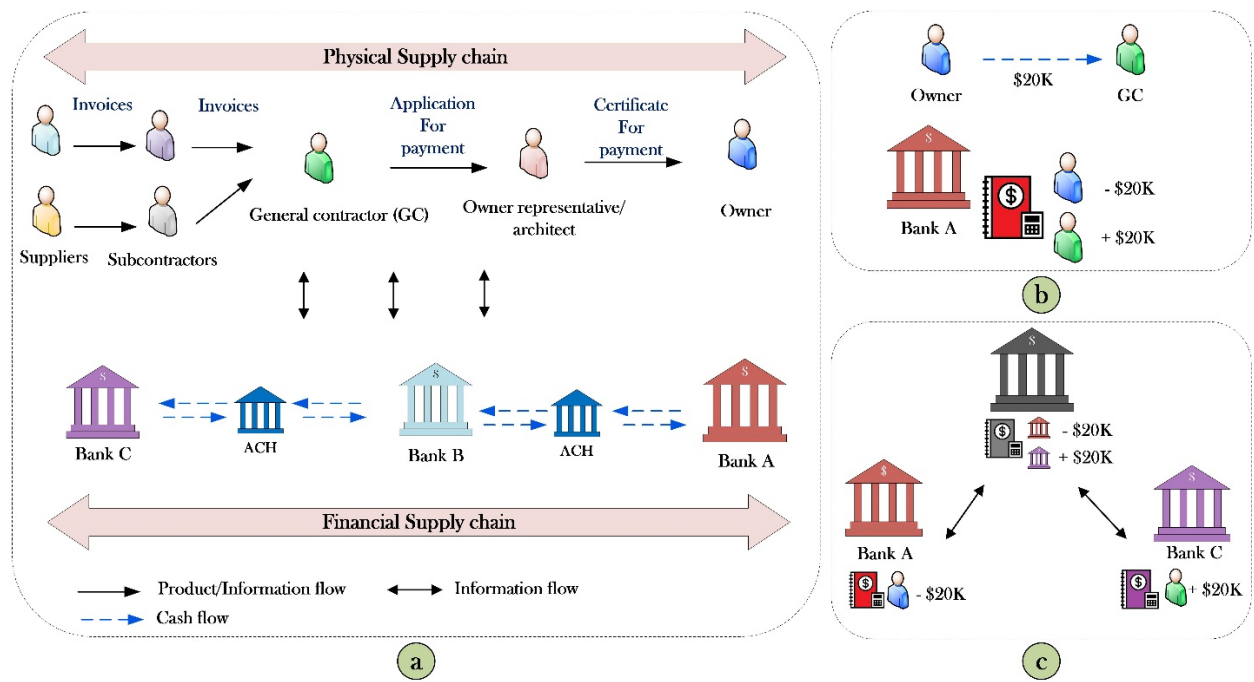

Fig. 1. The payment workflows in today's AEC industry: a) the lack of integration between physical and financial supply chains, b) payment settlement when parties hold accounts with the same bank or c) with different banks

In the more complex situations where stakeholders hold accounts with different banks (Fig. 1c). Correspondent banking is a model used in the past, where each bank holds accounts with other banks (Nostro account) and hosts accounts for them (Vostro account). Payments are settled by changes in the net liability of banks on the receiving and sending end of a transaction, reflected in their Vostro and Nostro accounts. More recently, payments between banks are handled using central banking models and batch processing approaches (Fig. 1c), where a network receives and batches transactions from the originating depository financial institutions (ODFI) and reports them to the receiving depository financial institutions

(RDFI); the resulting net debit and credit position of a bank is determined by the aggregate effect of outgoing and incoming transactions and is reflected in the bank's Nostro account with a central reserve (Fig. 1c). Examples of batch processing include the automated clearing house (ACH) (McAndrews 1994) in USA and the clearing house automated payment system (CHAPS) in UK (Becher et al. 2008).

In instances explained above, the bank *does not move money*; the payment is settled by *flowing information*, arithmetic operations on the banks' books, and not by flowing cash. This is a key factor that distinguishes current supply chains from those empowered by crypto asset use (Fig. 2):

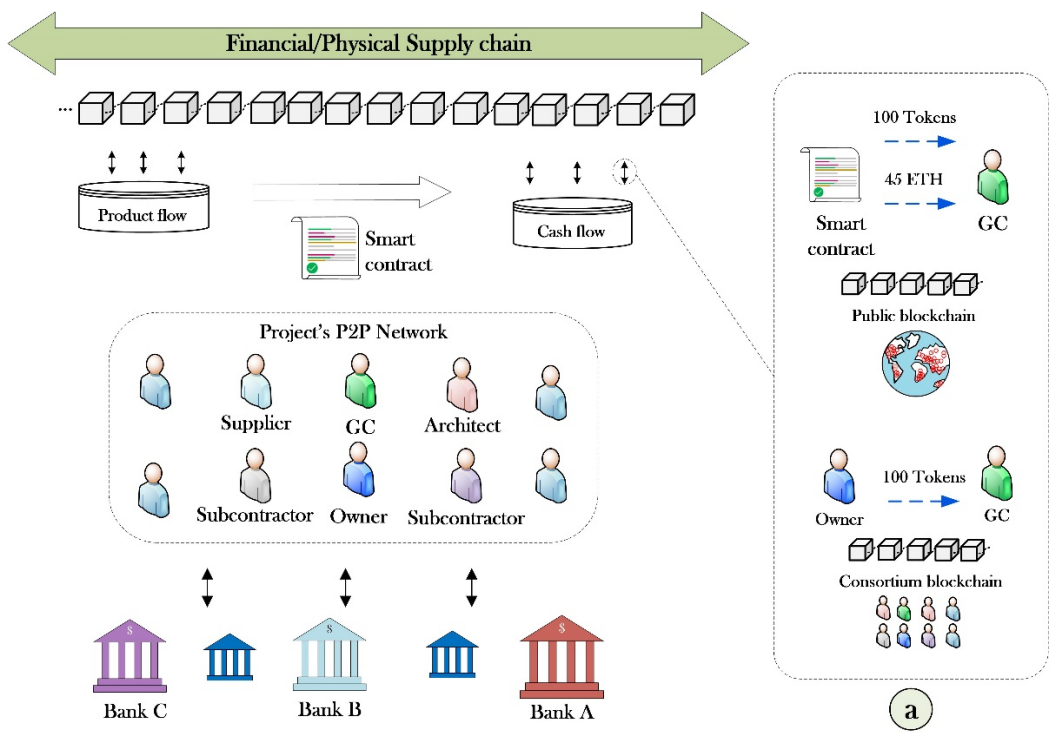

Fig. 2. Integration of physical and financial supply chains in the crypto asset-enabled supply chains

In a crypto asset-enabled system, the payment can be settled using native crypto currencies or overlay crypto tokens, with transactions formed between externally owned accounts (EOA), between contract accounts and EOAs, and between contract accounts; the first two are illustrated in Fig. 2a. Regardless of the crypto asset and the account type: 1) each account's balance represents the existence of actual reserves of crypto currencies, crypto tokens, or unique unspent transaction outputs; these funds are under the control of the party with access to the private key of the account address; 2) each payment changes the *ownership* of the aforementioned funds; this change represents a *movement of money*. This movement is written to and can be tracked on the blockchain. The payment moves crypto assets from one account to the next, such that the funds can be controlled only by the private key of the payee.

This *actual* movement of funds is what crypto asset-enabled supply chain in terms of integration; each movement of money provides an opportunity for integrating the product flow data with the payments (cash flow) they trigger. The updates in the status of product flows such as material deliveries, installations, and the construction of building elements at job sites can be directly tied to the crypto assets' change of ownership.

### 3.2 Two Facets of Crypto Asset-Enabled Integration: Granularity and Atomicity

The application of crypto assets for payments enhances 1) the granularity and 2) the atomicity of the integration between physical and financial supply chains:

#### 3.2.1 Granularity

This work characterizes granularity from three perspectives: 1) time, 2) trade, and 3) product. Therefore, a more granular integration between cash and product flows is herein characterized by payments that are more frequent, are payable to fewer parties, and are associated with smaller amount of work or fewer products (Fig. 3b).

- *Temporal granularity* refers to the period of *time* for which the construction work is documented in the payment applications and for which trades are entitled to a compensation. Increased temporal granularity equals more frequent payments.
- *Trade-level granularity* refers to the number of trades included in a payment. For example, a payment made to general contractor for indoor finishing work can include compensation for multiple trades such as insulation installer, drywaller, painter, and electrician. A more granular trade-level payment would involve fewer trades per transaction, resulting in more direct payments to each trade for their share of the work during a billing period.
- *Product-level granularity* refers to the number of products and building elements for which a party is compensated. For example, a payment to a general contractor for the concrete pour on one floor has higher product-level granularity than a payment for multiple floors.

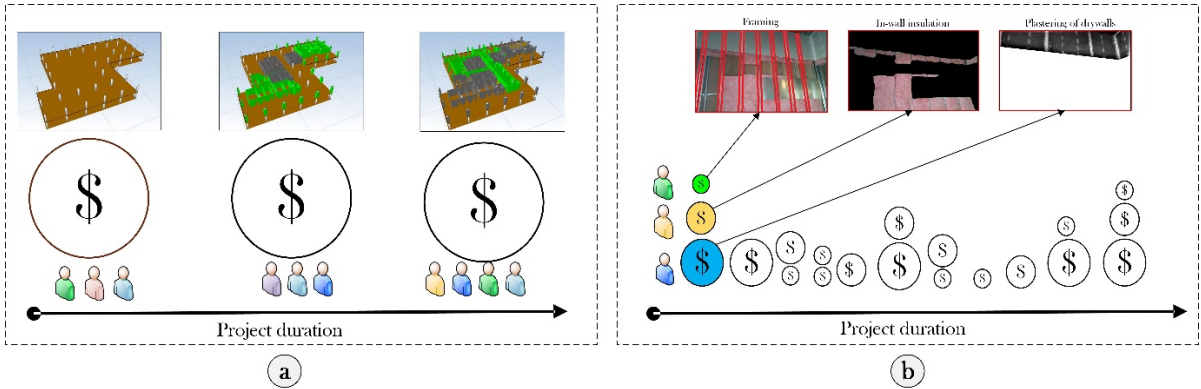

Fig. 3. A comparison between low and high granularity in payments: a) longer payment cycles, compensation to multiple trades, and for bigger scope of work; b) high temporal, trade-level, and product-level granularity

The increased granularity in the integration of cash and product flows makes it easier to align various sources of project data. The information, product, and cash flows are currently documented at different granularity and across sources such as 3D models, construction progress data, invoices, payment applications, and lien waivers, among others (Fig. 4). For example, the product documentation has the least granularity in the payment application submitted to owner, medium granularity in subcontractor invoices, and most granularity in the 3D models. These discrepancies make it difficult to connect these data sources; this problem is compounded when matching data using other features such as trade and time.

The time-consuming workflows associated with today's fragmented payment systems (Fig.1) hampers project stakeholders from achieving more granular integrations. Increased temporal, trade-level, and product-level granularity would result in exponentially higher number of payment applications and interactions with outside financial institutions; the resources required for preparation, review, approval, and the enforcement of these applications make such high granularity impossible to achieve. In the crypto asset-enabled system, on the other hand, the exchange of monetary value is decentralized and processed using an overlay protocol that benefits from the underlying blockchain's autonomy and consensus mechanism. This autonomy allows for direct use of product data for payment processing, increasing the granularity of integration. The third-party institutions are not involved in processing these granular payments; they instead are located at the periphery of the supply chain (Fig. 2), where they provide financial services to stakeholders.

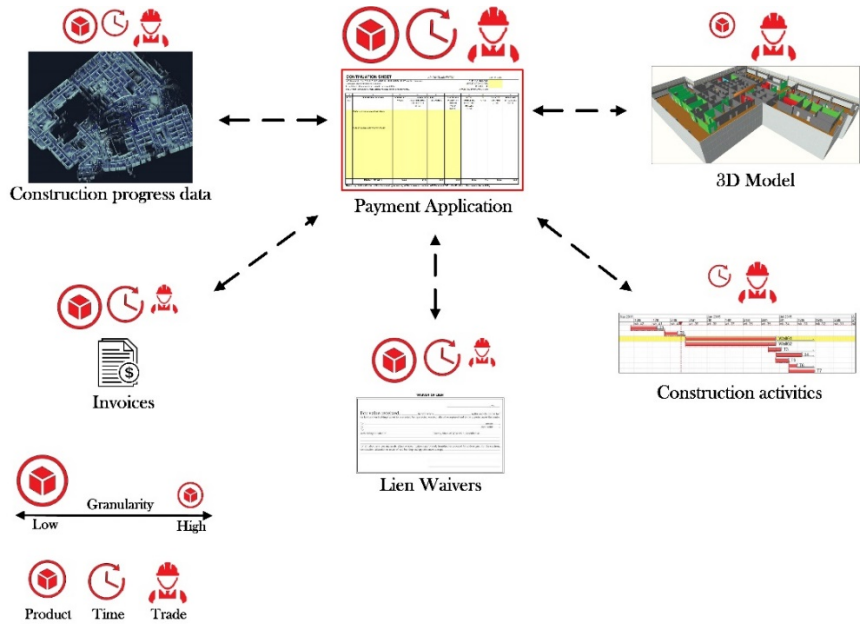

Fig. 4. The level of granularity significantly varies between different sources of project information

### 3.2.2 Atomicity

Atomicity is herein defined as both the product and cash flows documented within the same space; an atomic integration is characterized by a grouping of product flow status updates and the corresponding flow of cash into a single transaction. Increased atomicity contributes to the creation of a single source of truth and integrated information.

The construction supply chain currently lacks atomicity because the physical and financial supply chains are pushed apart from each other due to the heavily intermediated nature of the supply chain and reliance on external financial institutions (Fig. 1). This has resulted in a siloed data development ecosystem; the product flow is stored with project members and scattered through virtual models and remote captured progress data, whereas banks hold onto the corresponding cash flow data. The product flow status updates that trigger payments cannot be retrieved from the records of payments; the latter appear as a series of credit and debit operations on the bank's books and in aggregate form.

The application of crypto assets pushes financial institutions to the periphery (Fig. 2). This changes the role of banks in the construction supply chain, giving project stakeholders the ability to settle payments and agree on the concept of money without reliance on third-party intermediaries. As a result, the flow of funds is executed and documented within the supply chain network and not outside its boundaries. In this crypto asset-enabled supply chain, both the cash and product flows are documented in the same transactions, stored on the underlying blockchain (Fig. 2).

## 4  EXPERIMENT SET UP

A series of experiments were conducted in the context of construction progress payments to explore the role of crypto assets in enabling integration between product and cash flows. Fig. 5 illustrates the design of experiments, where the instances of crypto assets were used for payment processing. To validate this work's thesis, the resulting payment data was assessed in terms of the atomicity and granularity of integration between cash and product flows.

The product flow data (Fig. 5a) consists of two sets of 1) on-site reality captures, 2) as-built BIMs, and 3) artificial intelligence-enabled progress assessments for two commercial construction projects in California (USA) and Ontario (Canada). The first dataset (Law et al. 2020) is captured by an unmanned ground vehicle (UGV) equipped with a laser scanner, where three subcontractors were compensated for work on plumbing, indoor partitions, and heating, ventilations, and cooling. The second dataset (Hamledari et al. 2017a; Hamledari et al. 2017b) is captured by a camera-mounted unmanned aerial vehicle (UAV), where four subcontractors performed work on indoor partitions (framing, insulation, installation and plastering of drywalls, and painting).

The product flow data is passed to an autonomous smart contract-based payment method (Hamledari and Fischer 2020b; Hamledari et al. 2018) (Fig. 5b) which uses three instances of crypto assets: ERC721, ERC20, and ETH. The former is used for the transfer of lien rights; each token represents the right to physical property for which a payment is made. The ERC20 and ETH are used for on-chain payment settlement (see section 2.2 for an overview of these crypto asset classes). The smart contract-based solution stores product flow off-chain on a private IPFS network; the updates in the status of product flow are communicated to a smart contract on the Ethereum virtual machine (EVM) which performs the on-chain payment settlements and directly writes to the Ethereum blockchain.

Table 1. The two levels of granularity defined for product, temporal, and trade features of the payment data

|  | Level of granularity | |
|---|---|---|
|  | Low | High |
| Product | all elements of the same type | one building element |
| Time | monthly | weekly |
| Trade | general contractor | subcontractors |

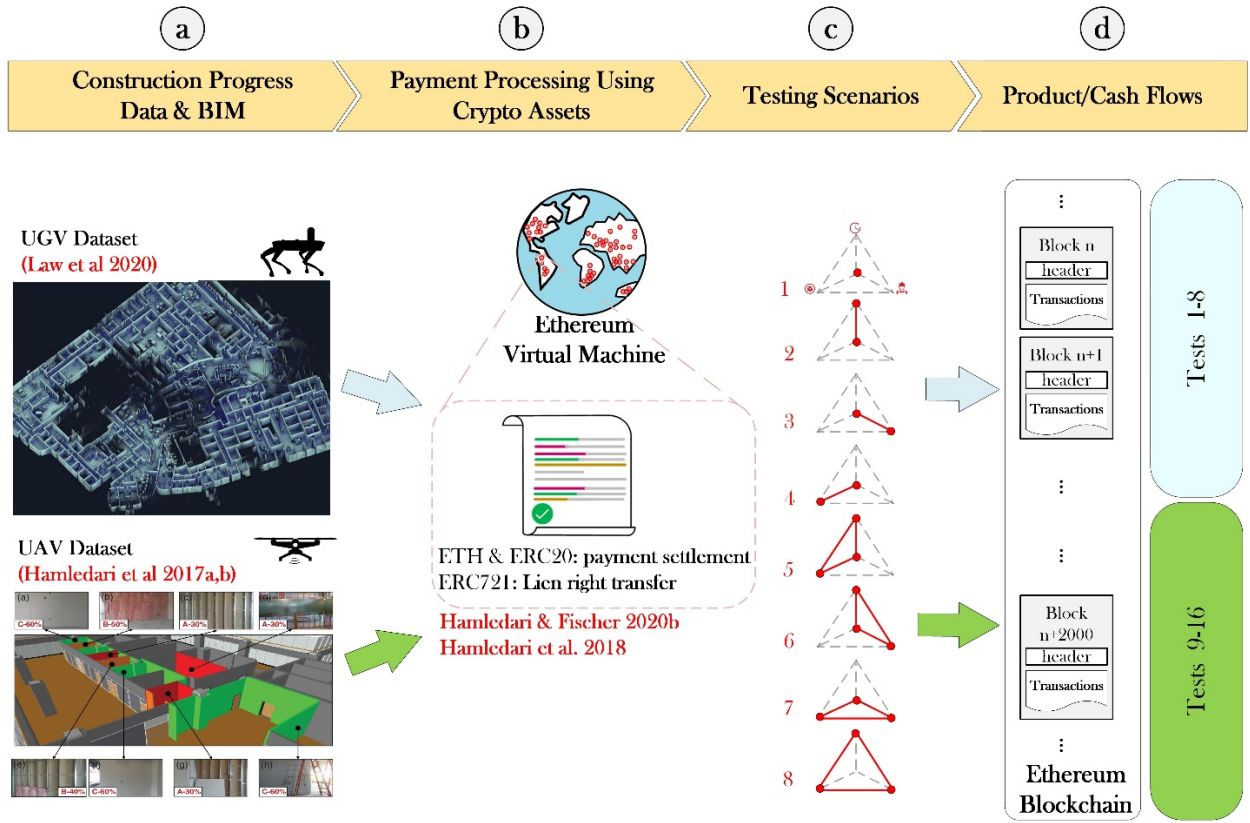

Fig. 5. The experiments designed to evaluate the effectiveness of crypto assets in enabling integration between cash and product flows: a) product flow data on two commercial construction projects, 2) smart contract-based method used for payment processing, c) the eight testing scenarios defined based on variations in the product-level, temporal, and trade-level granularity, and d) the resulting 18 datasets of payment data

To evaluate the effectiveness of crypto assets in enabling granularity, the low and the high level of granularity were defined for product-level, temporal, and trade-level features of the payment data (Table 1). For example, the low and high trade-level granularity respectively correspond to payments from the owner to general contractor (GC) and the direct payments from the owner to subcontractors.

Fig. 6 illustrates these levels of granularity for product-level, temporal, and trade-level features. The combination of the low and high granularity for these three features resulted in a total of 8 testing scenarios (Fig. 5c). Each of the UGV and UAV datasets was used to generate 8 sets of payment data, ranging from the lowest granularity (i.e., monthly payment from the owner to GC for work on all building elements) to highest (i.e., weekly payment from the owner to a subcontractor for work on one particular building element). These tests were run in parallel and for a period of one month.

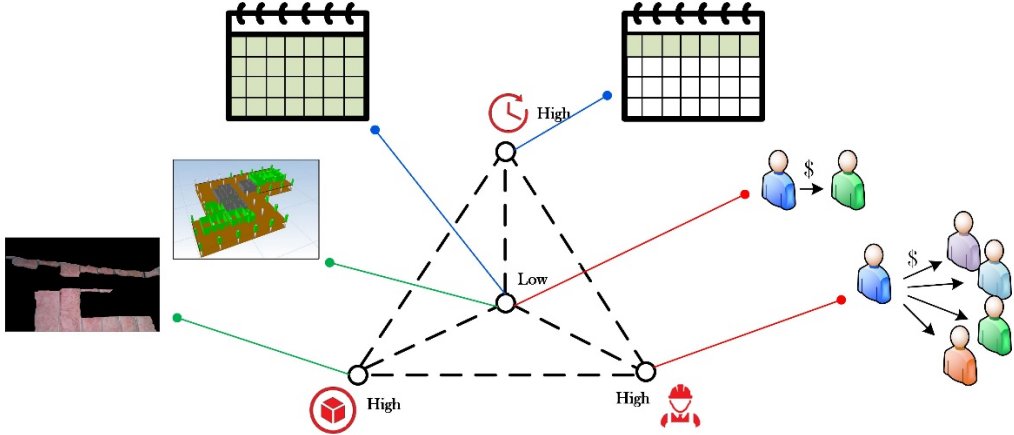

Fig. 6. The low and the high levels of granularity for product, time, and trade

## 5 RESULTS & DISCUSSION

The experiments resulted in 16 sets of payment data (Fig. 5d), consisting of transactions written to the blockchain. Tests 1-8 only vary in terms of their granularity and otherwise correspond to the same scope of work on the first project. This is also the case for tests 9-16, corresponding to payments on the second project.

The payment datasets were examined to verify whether *granularity* and *atomicity* were achieved in the integration between cash and product flows; the results are detailed in sections 5.1 and 5.2 respectively. The implications of these findings are further discussed in section 5.3.

### 5.1 Granularity of Integration

The integration is successfully achieved if the payments have the intended level of granularity as specified in the testing scenarios (Fig. 5c). According to the on-chain transactions for the UGV dataset (Fig. 5a), payments in the tests 4, 5, 7 and 8 failed to materialize; these unsuccessful tests all correspond to high product-level granularity. As for the UAV dataset (Fig. 5a), all eight tests (9-16) were successfully executed and achieved the intended granularity.

Further analysis of the tests' input data reveals that this discrepancy is due to the differences between the product flow's level of detail (LoD) in the two datasets. In both projects, the BIM provides design and construction data for each building element, available per globally unique identifiers (GUID). The progress assessments (i.e., as-built condition at the jobsites), on the other hand, are provided in aggregate form for the UGV dataset and per GUID in the UAV dataset. As a result, the high product-level granularity failed to materialize in payments associated with the former.

Fig. 7 illustrates four transactions achieved using ETH and ERC20 for payment settlement. These transactions vary in terms of their levels of granularity and are respectively part of the transaction pools in tests 1 (low granularity), 4 (high product granularity), 8 (high product, temporal, and trade granularity), and 6 (high temporal and trade granularity). They respectively correspond to 1) payment to the GC for the framing, insulation, drywall installation, plastering, and painting performed during the month of June on 104 partitions (Fig. 7a); 2) payment to the GC for the framing, insulation, drywall installation, and plastering performed on one particular partition during the week of June 4-11 (Fig. 7b); c) payment to the framing subcontractor for the work on one particular partition during the week of June 4-11 (Fig. 7c); and d) payment to the insulation subcontractor for the insulation work performed on 42 partitions during the week of June 12-19.

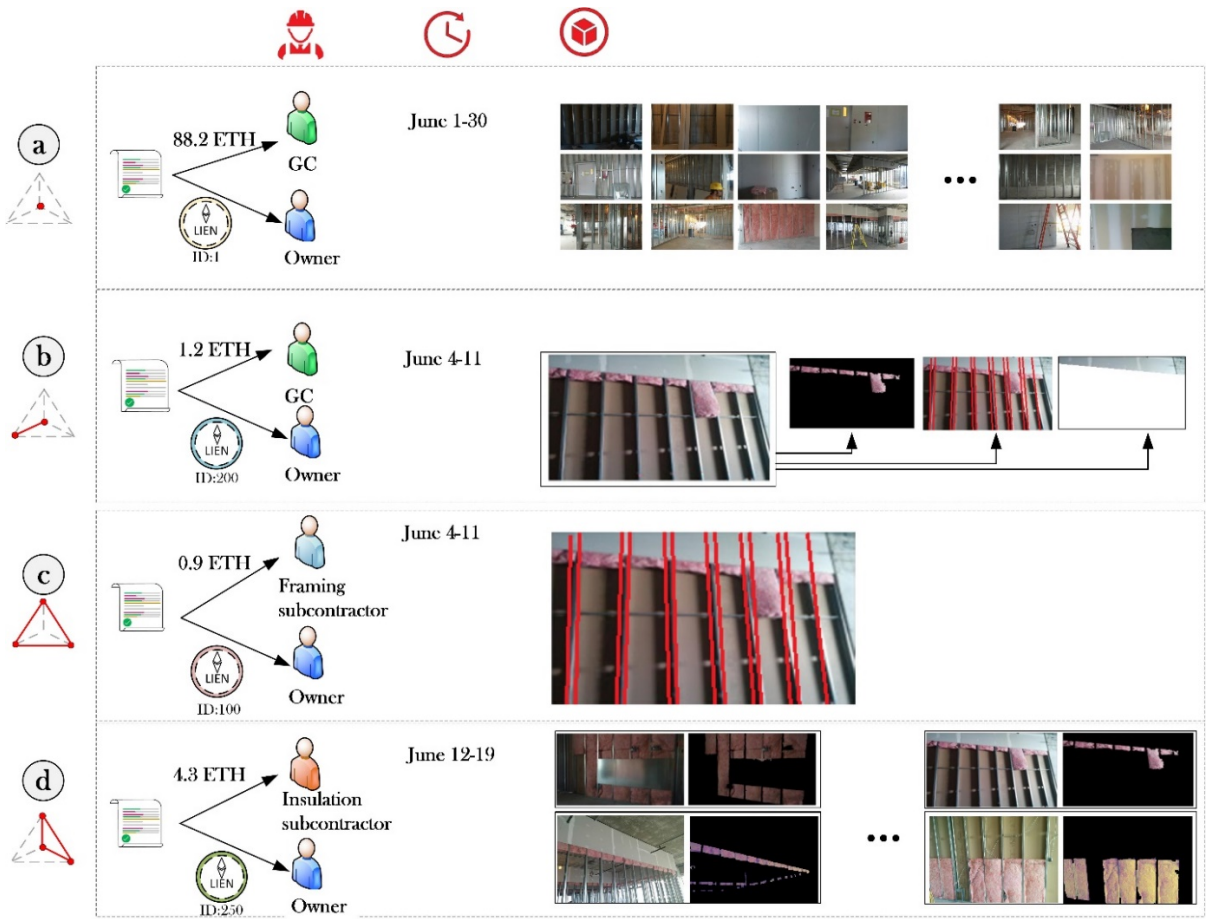

Fig. 7. Different levels of granularity achieved using ERC20 and ETH in payment settlement: a) low granularity, b) high product granularity, c) high product, temporal, and trade granularity, and d) high temporal and trade granularity

This increased granularity is not limited to the payment settlement and is extended to the transfer of lien right. As shown in Fig. 7, the smart contract-based method (Hamledari and Fischer 2020b) uses the ERC721-based LIEN token to transfer the lien right to the owner for the scope of work for which the trades are compensated in the ERC20/ETH transaction. These two transactions are broadcasted simultaneously by the smart contract. As the construction progresses, instances of LIEN token are minted and transferred to the owner, with each having its own unique ID. Fig. 8 shows the LIEN token corresponding to the payment depicted in Fig. 7c; it represents the right to the framing work on the partition shown in Fig. 8b.

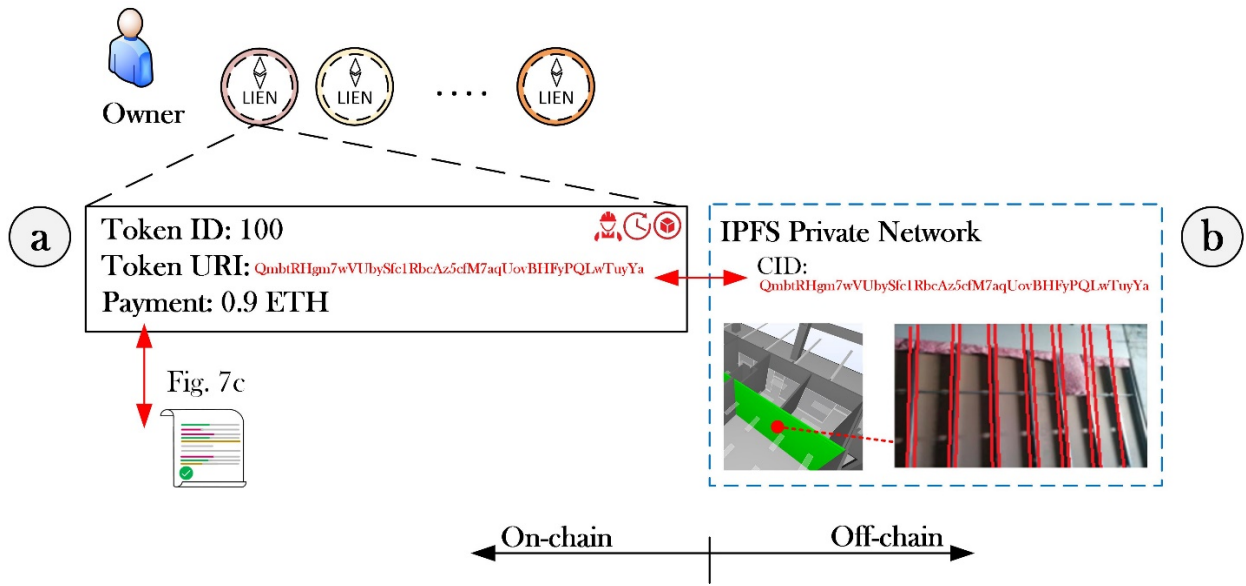

Fig. 8. The application of ERC721 in the transfer of lien rights: granular definition of scope of work in the LIEN token and the atomic integration between product data (off-chain) and financial data (on-chain)

## 5.2 Atomicity of Integration

The transactions in all tests were observed to achieve atomicity in their integration of cash and product flows: the product flow updates corresponding to each flow of cash were directly retrieved from the payment transactions and vice versa. These flows are stored within the same space (Fig. 9) and not in separate siloes.

Fig. 9 illustrates how the application of the crypto assets enables this atomic integration: the EVM is a singleton state machine, and the transactions processed by its miners result in changes to the state of the Ethereum blockchain. This change is reflected in the *state trie* (also "state tree") (Fig. 9c). It is a modified Merkle Patricia trie, a global tree-like data structure that maps key-value pairs, where the key and value are respectively an account address and the encoding of its *nonce*, *balance*, *StorageRoot*, and *CodeHash* (Fig. 9d). These changes are recorded for both EOAs and contract accounts.

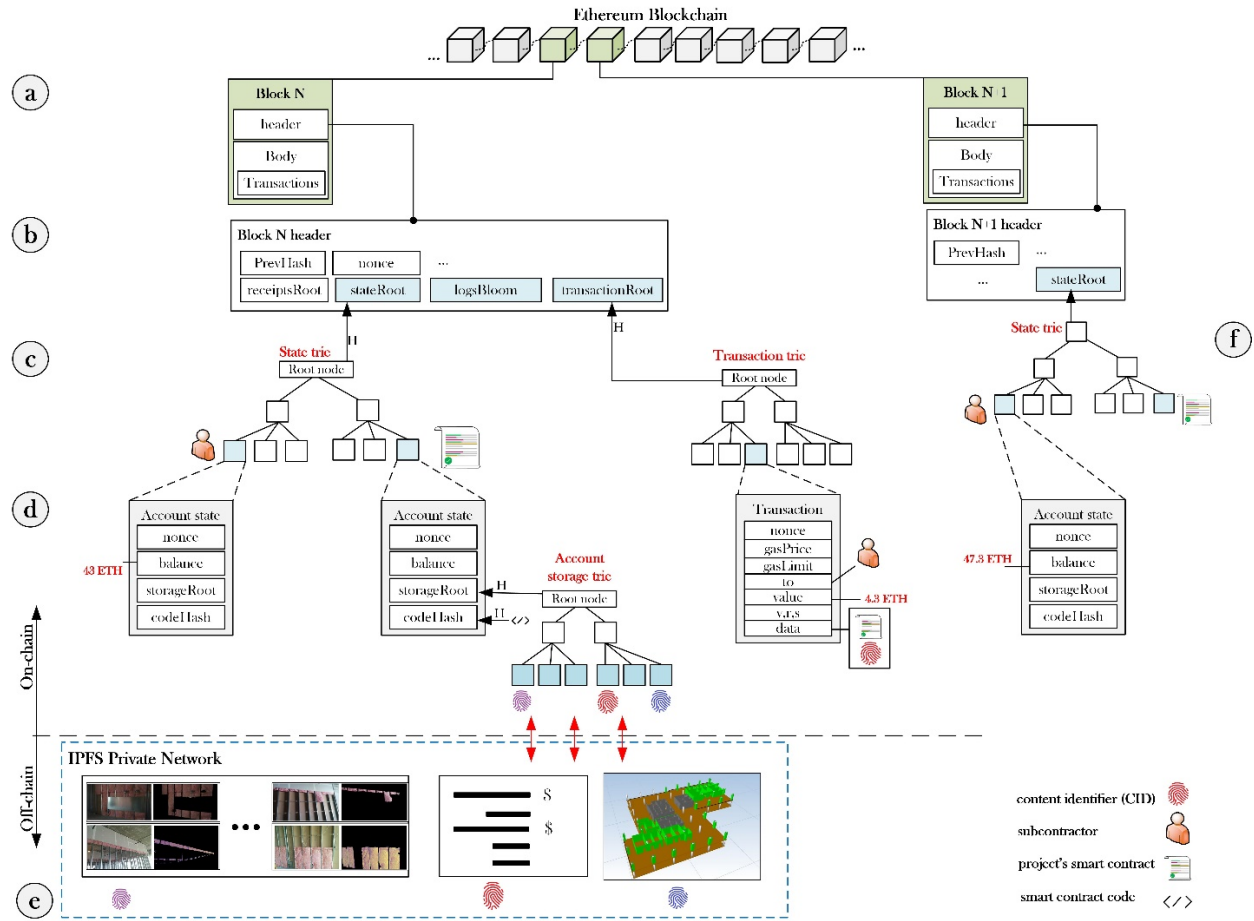

Fig. 9. Crypto asset application enables atomic integration of cash and product flows, connecting the off- and on-chain realities

The *StorageRoot* is the 256-bit hash of the *account storage trie*, where all the smart contract data and the mapping between product and cash flow is stored. This includes the content identifiers (CID) corresponding to the product flow updates stored off-chain and on the project's IPFS private network (Fig. 9e); the off-chain data include, among others, the construction progress, as-built BIM, and the schedule of values. Each transaction, stored in the *transaction trie* and the body of a block (Fig. 9a), can be linked with the corresponding CID that initiated that transaction, creating an atomic integration where both flows are stored within the same space. The CIDs cannot be interpreted by individuals outside the IPFS network and hence do not jeopardize the privacy of the data.

As shown in Fig. 8, this atomicity is also extended to the transfer of lien rights. As part of the ERC721 standard, each LIEN token has a uniform resource identifier (URI) which stores its metadata. A LIEN token's URI stores the CID of the product flow updates for which the right is transferred. For example, the token shown in Fig. 8a stores the CID referencing the framing work on the partition shown in Fig. 8b. This atomicity makes it possible to retrieve the monetary value and the product flow associated with each transfer of lien right *directly* from the LIEN token.

### 5.3 Implications for the AEC Industry

The application of crypto assets was observed to enhance the atomicity and the granularity of the integration between cash and product flows. In contrast with today's payment systems, the crypto asset use provides a means of payment processing and the transfer of lien rights for the scopes of work that have varying levels of granularity. This enables the flow of funds on projects to occur more frequently, to compensate smaller scopes of construction work, and to be more directly payable to the trades.

For instance, the application of ETH and ERC20 made it possible for a project to compensate a subcontractor on a weekly basis for the framing work on a single interior partition. The ERC721 use allowed for the transfer of lien right corresponding to the same granular scope of work. These transactions were recorded in an immutable manner and timestamped on the Ethereum blockchain, creating a permanent and auditable link between the flow of funds and the corresponding product flow updates. The record of each payment provides access to the construction progress data triggering that payment along with project information such as BIM, schedule of values, and the data analytics tools used in the valuation of the work. This level of granularity and atomicity is not possible in today's supply chains.

Based on the results, it is evident that the product flow's LoD plays a critical role in achieving increased granularity. The use of crypto asset alone does not guarantee an increase in granularity. This incentivizes the development of robust product flow management systems using robotics and artificial intelligence. Stakeholders that invest in such data-driven solutions across their supply chain can benefit from an increased granularity in their financial supply chain, enabled by crypto asset use. These same investments do not currently translate to enhancements in the payment systems.

The payment applications currently used in the industry have the same granularity regardless of how technology is used in managing the physical supply chain; the end result is blind to improvements in the on-site use of reality capture technologies, the robust tracking of construction material, and the use of as-built BIMs. In the crypto asset-enabled system, on the other hand, the financial supply chain can be as granular as the physical supply chain, and so can their integration. Increases in the frequency of data capture, the LoD of progress data, and the detailed modeling of trades in BIM directly translate to enhancements in payment's granularity.

Achieving atomicity, on the other hand, appeared to be independent of the product flow's LoD. The application of crypto asset increases atomicity even when the payment granularity is as its lowest and matches that of today's payment applications. Both the physical and financial supply chain data can be retrieved directly from the on-chain transactions or the instances of the crypto assets. The stakeholders do not need to run full nodes in the public blockchain to access this data.

Integrating information is key to successful project delivery and creating a single source of truth (Fischer et al. 2017). The combination of atomicity and granularity, enabled by crypto asset use, can enhance the alignment between project data stored across different data sources (Fig. 4); this is because information can be stored in a single space, rather than in siloes, and it can be highly granular, allowing for successful matching across different sources such as BIM, activities, schedule of values, and on-site observations. Increased transparency is another potential impact of such improvements. In the blockchain data structures such as state trie, storage trie, and transaction trie, the changes propagate upwards; any alteration of the project data results in inconsistencies that can be automatically detected. This has potential to both detect and prevent fraud, one of the common forms of corruption in the construction industry (Le et al. 2014).

## 6  LIMITATIONS & RISKS

The experiments conducted in this work used datasets that were collected at different construction projects, described work conducted on a variety of building elements, and were collected using different sensors. However, they both focused on the on-site installation and construction of elements. They did not, for example, include off-site construction and the delivery of materials. That said, the on- and off-chain documentation of flows occurs independent of the type of product flow status updates, the type of sensor, and the source of the product flow data. Therefore, it is expected that crypto asset use enables the same integration for those scopes of work. Future work should validate this by extending the experiments to a broader range of activities.

In this work, the experiments were conducted on the Ethereum blockchain, a permission-less chain. While it is expected that the findings extend to private or consortium blockchains, this was not validated. In private chains, the integrity of the transactions depends on the design of the consensus algorithm and the possibility of collusion between parties. It is not certain that the integration of physical and financial supply chain, once achieved on a private chain, will remain immutable and secure. On the other hand, the application of permission-less chains in this work necessitates a closer look at privacy concerns; this was not part of the scope and needs to be addressed in future.

The application of crypto assets for conditioning the cash flow on product flow may expose projects to risks associated with 1) the regulations and 2) the price volatility of crypto currencies:

The crypto asset space is fast evolving, and so are the regulations surrounding their use. Crypto assets may be considered securities (De Filippi and Wright 2018); this can make them subject to scrutiny by regulatory bodies such as the Securities and Exchange Commission (SEC) and the Federal Trade Commission (FTC). While there is debate as to whether more regulation is needed, some case studies suggest that the current regulations are sufficient in addressing the risks of crypto assets to financial stability and monetary policy (Manaa et al. 2019). The regulatory bodies need to provide clarity as to which tokens are considered security (Edwards et al. 2019). This reduces the uncertainties for the industry applications.

Crypto currencies have higher price volatility compared with fiat currency. While stablecoins (Calcaterra et al. 2019) can offer a feasible alternative, the increased adoption of crypto currencies can reduce the volatility in future. The price of a crypto currency is a function of its utilization and the market's speculation about its potential value as an investment asset. As more industries adopt a crypto currency, its utilization increases, and the changes in the speculative component will have smaller influence on the overall price of that crypto currency. Increased adoption also means the availability of more buyers and sellers; this increases market liquidity, further reducing the price volatility.

# 7 CONCLUSION

In response to recent technological disruptions, the construction and engineering industry is projected to undergo consolidations across its value chain; this can particularly benefit the financial supply chain and its integration with the physical supply chain. The transition from product flow to payments is currently heavily intermediated; the product flow is scattered across several supply chain partners, and the cash flow sits with external financial institutions. There is a disconnect between the two.

This work demonstrated how the application of blockchain-based crypto assets (crypto currencies and crypto tokens) can enable integration when used in the place of fiat currency and for conditioning the payments on the flow of products. The thesis was validated in a series of experiments where crypto assets were used for processing payments to subcontractors on two commercial construction projects. The findings indicate that supply chains empowered by crypto assets can benefit from increased granularity and atomicity in their integration of flows. The increase in the granularity, however, is dependent on the product flow's LoD. This motivates further attention automated and data-driven approaches toward the management of product flow. The atomicity, on the other hand, can be achieved regardless of the LoD for product flow. Projects can benefit from crypto asset-enabled integration regardless of the LoD used in their product flow management. The investments in the latter, however, can directly translate to improvements in the financial supply chain.

While this work examined the mechanisms contributing to the integration of flows, it is crucial to analyze the impact of such integration on the supply chain visibility and the performance of supply chain partners. It is not currently clear how enhancements in the granularity and atomicity of integration affects the experience of stakeholders retrieving and using the data. This necessitates the design of experiments where the accuracy of the data retrieval, the level of effort, and other measures of supply chain visibility can be quantified in supply chains using fiat currency and crypto assets.

## 8  ACKNOWLEDGEMENT

This work is financially supported by the Center for Integrated Facility Engineering (CIFE) at Stanford University (grants 2020-09, 2018-06, 2017-06). The authors are grateful to Swinerton, PMX Construction, Perkins+Will, Inc., for their support during the data collection phase and granting access to project data. The first author extends his gratitude to Eric Law and Tristen Magallanes (Swinerton); Dr. Kincho Law, Dr. Michael Lepech, Dr. Forest Flager, Alissa Cooperman, Parisa Nikkhoo, and Tulika Majumdar (Stanford University); Dr. Brenda McCabe, and Pouya Zangeneh (University of Toronto) for their immense role in his PhD journey and for their invaluable intellectual companionship.